\documentclass{elsart}

\usepackage{epsfig}

\usepackage{amssymb}

\begin{document}

\begin{frontmatter}



\title{Phase-field modeling of the discontinuous precipitation reaction}


\author[Alger,Palaiseau]{Lynda Amirouche},
\author[Palaiseau]{Mathis Plapp}

\address[Alger]{Laboratoire de Physique Th\'eorique, Facult\'e de Physique,
U. S. T. H. B., BP 32, El-Alia, BabEzzouar, 16311, Alger, Algeria}
\address[Palaiseau]{Physique de la Mati\`ere Condens\'ee, 
\'Ecole Polytechnique, CNRS, 91128 Palaiseau, France}

\begin{abstract}
A multi-phase-field model for the description of the discontinuous 
precipitation reaction is formulated which takes into account
surface diffusion along grain boundaries and interfaces as well
as volume diffusion. Simulations reveal that the structure and 
steady-state growth velocity of spatially periodic precipitation 
fronts strongly depend on the relative magnitudes of the
diffusion coefficients. Steady-state solutions always exist for a 
range of interlamellar spacings that is limited by a fold singularity
for low spacings, and by the onset of tip-splitting or oscillatory
instabilities for large spacings. A detailed analysis of the
simulation data reveals that the hypothesis of local
equilibrium at interfaces, used in previous theories, is not 
valid for the typical conditions of discontinuous precipitation. 
\end{abstract}

\end{frontmatter}

\def\be{\begin{equation}}
\def\ee{\end{equation}}
\def\mother{{\alpha_0}}
\def\dbulk{D_v}
\def\dsurfma{D_b}
\def\dsurfmb{D_b^{\mother\beta}}
\def\dsurfab{D_b^{\alpha\beta}}

%
%

\section{Introduction}

The discontinuous precipitation reaction is a solid-state
transformation during which a supersaturated mother phase $\mother$
decomposes into a two-phase structure consisting of the 
depleted $\alpha$ phase and lamellar precipitates of a new phase $\beta$.
This reaction takes place at a moving grain boundary, which
indicates that the rate-limiting step is the diffusion of 
solute along grain boundaries and interfaces. The resulting
characteristic lamellar microstructure has been observed in 
a large number of different alloy systems \cite{Manna01}.

Numerous theories have been proposed to predict the precipitate
growth velocity and the interlamellar spacing as a function 
of the processing conditions and the alloy 
thermodynamics \cite{Cahn59,Klinger97,Brener99} 
(see also \cite{Manna01} and references therein), 
but the direct comparison of these predictions to
experimental data is made difficult by the fact that the 
process is extremely complex and controlled by a large number 
of parameters which are often not precisely known. In this
situation, direct numerical simulations of the discontinuous 
precipitation reaction can help to reach a better understanding 
of this phenomenon: if a reasonably realistic model can be 
investigated in detail, the simulation data can be used to
test the theories and to clarify whether their basic assumptions
are valid.

The phase-field method, which is by now a well-established 
simulation tool both in crystal growth and phase 
transformations \cite{Chen02,Boettinger02}, is ideally suited
for this purpose. In phase-field models, the local
state of matter is described by one or several order
parameters (the phase fields), and interfaces
are represented by rapid variations of these fields over
a characteristic length scale $W$. By choosing $W$ small enough, 
all the details on the length scale of the microstructure 
can be properly treated. The evolution equations for the fields are 
obtained from the principles of out-of-equilibrium thermodynamics, 
and therefore only a small number of assumptions is needed to
obtain a fully consistent model for discontinuous precipitation.
Whereas a closely related phenomenon, the so-called discontinuous
spinodal decomposition that takes place at moving grain boundaries,
has recently been investigated by a phase-field 
model \cite{Ramanarayan04}, to our best knowledge no previous
phase-field study of discontinuous precipitation is available.

Here, we develop a phase-field model for the discontinuous 
precipitation reaction by modifying a recent 
model for eutectic solidification \cite{Folch03,Folch05}.
We use the multi-phase-field approach
\cite{Steinbach96,Nestler05,Eiken06}, in which each phase
(or grain) is described by one phase field. We restrict 
our attention to isothermal growth in a simple binary alloy; 
extensions of the model to more complicated situations 
are straightforward.

We then carry out simulations and vary systematically the
parameters of the model to investigate under which conditions
steady-state growth of spatially periodic lamellar precipitate
arrays is possible. In particular, we study the influence 
of lamellar spacing, alloy composition, and interfacial 
parameters (surface tensions and surface diffusivities)
on the growth velocity. As a guideline, we use insights
from a detailed recent sharp-interface model developed by 
Brener and Temkin \cite{Brener99}. We confirm qualitatively
several features predicted by this model, in particular 
the importance of the contact angles between the interfaces at 
the trijunction point and the decisive role of solute diffusion 
along the interphase boundaries {\em behind} the growth front.
However, the values of the precipitate growth velocity obtained
from our simulations do not agree with the theoretical predictions,
and the shape of the velocity-vs-spacing curves obtained here
strongly differs from the ones obtained in Ref.~\cite{Brener99}.
A detailed analysis of the simulation data reveals the main reason 
for this discrepancy: the local equilibrium hypothesis used in the
sharp-interface models is not valid for diffuse interphase boundaries
in the presence of strong surface diffusion. We believe that this
effect is generic, and we develop a criterion that clarifies under
which conditions it becomes important.

We also observe a new instability that occurs for large spacings 
and leads to oscillatory growth, reminiscent of the oscillatory
patterns found in eutectic solidification \cite{Karma96,Ginibre97}. 
This could be related to the ``stop and go'' motion of discontinuous 
precipitation cells observed in several alloy systems \cite{Manna01}.

The remainder of this paper is organized as follows: in
Sec.~\ref{sec:model}, we present the phase-field model 
and relate its parameters to the ones used in conventional 
sharp-interface theories. In Sec.~\ref{sec:simulation}, we 
give a few details about our simulation procedures. Results 
are presented in Sec.~\ref{sec:results}, followed by a 
discussion in Sec.~\ref{sec:discussion} and a conclusion 
in Sec.~\ref{sec:conclusion}.

\section{Model}
\label{sec:model}

\subsection{Phase-field formulation}

We seek to construct a model that can reproduce the
phenomenon of discontinuous precipitation but remains
as simple as possible. Therefore, we will make a number 
of simplifying assumptions:
\begin{enumerate}
\item We consider isothermal processing of a binary A-B alloy 
and suppose that the lattice constant is independent of 
composition. This allows us to disregard elastic effects. 
As a consequence, both the thermodynamics and the kinetics 
of the model are governed by the composition field only.
\item We assume that there is no grain boundary segregation. 
Then, grain boundaries exhibit motion by curvature only
(no solute drag effects).
\item We choose a particularly simple alloy thermodynamics
by assuming that the free energy densities of the two phases
involved ($\alpha$ and $\beta$) are simple parabolas of
equal curvature.
\end{enumerate}
All of these assumptions could be relaxed by constructing a 
more general phase-field model along the lines of previous 
works \cite{Chen02}, but this would considerably complicate
the analysis.

Our approach is based on a phase-field model for
two-phase solidification that has been presented in 
detail in Refs.~\cite{Folch03,Folch05}. Each of the 
three involved phases -- the mother phase $\mother$, 
the depleted $\alpha$ phase and the precipitate phase $\beta$ --
is described by one phase field $p_i$ which represents 
the local volume fraction of the corresponding phase. 
The phase fields are hence constrained by the condition
\be
p_\mother + p_\alpha + p_\beta = 1
\ee
for all space points. 

The alloy thermodynamics is described by free energy densities 
$f_\alpha(C)$ and $f_\beta(C)$, where $C$ is the alloy composition
(atomic fraction of B atoms). The mother phase $\mother$ and the
depleted $\alpha$ phase have different crystallographic orientations
and are separated by a grain boundary, but are thermodynamically
identical. For the fixed processing temperature, two-phase
equilibrium is characterized by the equilibrium compositions
$C_\alpha$ and $C_\beta$ and the equilibrium chemical potential
$\mu_{\mathrm eq}$. For simplicity, we make the transformation 
$\mu\to \mu - \mu_{\mathrm eq}$ and $f_i\to f_i-\mu_{\mathrm eq}C$ 
$(i=\alpha,\beta)$, which shifts the equilibrium chemical 
potential to zero. Next, we assume that the free energy densities
are parabolic around the equilibrium compositions,
\be
f_\alpha = \frac{A}{2} (C-C_\alpha)^2 \quad {\rm and} \quad
f_\beta = \frac{B}{2} (C-C_\beta)^2,
\ee
where $A$ and $B$ are constants of dimension energy per unit volume.
It turns out that the construction of the model is particularly
simple if we set $A=B$, which means that the two parabolas have 
equal curvatures. Since we are interested here in generic features
of discontinuous precipitation, there is no harm in making this
choice. We then define a scaled composition $c$ by
\be
c = \frac{C-C_\alpha}{C_\beta-C_\alpha}.
\ee
The free energy densities, expressed in this variable, are then
\be
f_\alpha(c) = \frac 12 H_c c^2 \quad {\rm and}\quad
f_\beta(c) = \frac 12 H_c (c-1)^2,
\ee
where $H_c=A(C_\beta-C_\alpha)^2$.

The starting point for defining the dynamics is a free
energy functional which depends on the phase fields and
the concentration field,
\be
F[{\bf p},c] = \int \frac 12 K \sum_i |\vec\nabla p_i|^2 + 
H_p f_{TW}({\bf p}) + \frac 12 H_c [c - g({\bf p})]^2,
\label{ffunctional}
\ee
where $K$ and $H_p$ are constants of dimension energy per
length and energy per volume, respectively, 
${\bf p} \equiv \lbrace p_\alpha, p_\mother, p_\beta\rbrace$
is the set of phase fields, the sum runs over all phases
($i=\mother,\alpha,\beta$), and $f_{TW}$ and $g$ are
dimensionless functions that depend only on the phase fields.
The former, $f_{TW}$, creates a potential landscape for the
phase fields with three distinct minima, corresponding to
the three phases (for phase $i$, $p_i=1$ and the other
phase fields are zero). Its expression is
\be
f_{TW} = \sum_i p_i^2 \left(1-p_i\right)^2 + 
a_\alpha p_\mother^2p_\beta^2
   \left(2p_\mother p_\beta+3p_\alpha+6p_\alpha^2\right).
\ee
With $a_\alpha=0$, the potential is symmetric with respect to
the exchange of any two phases, which implies that the
surface tensions of all interfaces are equal. The term
proportional to $a_\alpha$ breaks this symmetry and modifies
the surface tension of the $\mother$-$\beta$ interface
without modifying the others (see below for more details).
The function 
\be
g({\bf p}) = \frac 14 p_\beta^2 \left\lbrace 15(1-p_\beta)
\left[1+p_\beta-\left(p_\mother-p_\alpha\right)^2\right] +
p_\beta\left(9p_\beta^2-5\right)\right\rbrace
\ee
couples the phase fields to the scaled composition. It satisfies
$g(p_\beta=1)=1$ and $g(p_\beta=0)=0$. Therefore, the last 
term in the free energy functional is identical to the
free energy of the $\alpha$ and $\beta$ phases for 
$p_\beta=0$ and $p_\beta=1$, respectively.

It is convenient to introduce a dimensionless free energy
functional by dividing the free energy density by the 
constant $H_p$, which yields
\be
{\cal F} = \int \frac 12 W^2 \sum_i |\vec\nabla p_i|^2 + 
f_{TW}({\bf p}) + \frac 12 \tilde\lambda [c - g({\bf p})]^2,
\label{ffunctionalnondim}
\ee
where $W=\sqrt{K/H_p}$ is the characteristic length scale
of the diffuse interfaces, and $\tilde\lambda=H_c/H_p$
is the ratio of the energy scales associated with the
phase-field and concentration contributions in the free energy.

The evolution of the phase fields and the concentration
field is obtained from this free energy functional by
variational derivatives. We have
\begin{equation}
\tau \frac{\partial p_i}{\partial t} =
        - \left.\frac{\delta \cal F}{\delta p_i}
        \right|_{\sum_i p_i = 1},
\end{equation}
where $\tau$ is the relaxation time for the phase 
fields. The functional derivative on the right hand side 
has to be evaluated taking into account the constraint on the 
phase fields, which can be done using a Lagrange multiplier
as detailed in Ref.~\cite{Folch05}. For the concentration 
field, we have the standard conservation law,
\be
\partial_t c = \vec\nabla\cdot
   \left(M({\bf p})\vec\nabla\frac{\delta{\cal F}}{\delta c}\right),
\label{cequation}
\ee
where $\delta{\cal F}/\delta c\equiv\mu$ is the chemical potential, 
and $M({\bf p})$ is the mobility of the solute. The latter is 
written as
\be
M({\bf p}) = D({\bf p})/\tilde\lambda,
\ee
where $D({\bf p})$ is the local diffusivity. It is easy to
verify from Eqs. (\ref{ffunctionalnondim}) and (\ref{cequation})
that this choice yields Fick's law in the bulk. For simplicity,
we assume that the volume diffusion coefficient $\dbulk$ is the
same for the $\alpha$ and $\beta$ phases. In contrast, it
is important to include different surface diffusion 
coefficients for each surface. We define
\be
D({\bf p}) = \dbulk + 4p_\mother p_\alpha \dsurfma + 
                      4p_\mother p_\beta \dsurfmb +
                      4p_\alpha p_\beta \dsurfab,
\ee
where $\dsurfma$ is the grain boundary diffusion coefficient and
$\dsurfmb$ and $\dsurfab$ are the surface diffusion coefficients
for the interphase boundaries between the precipitate and the
supersaturated and depleted $\alpha$ phases, respectively.
This form of the surface diffusivity terms is motivated by 
the fact that the product $p_ip_j$ is zero in the bulk phases 
and has a maximum value of $1/4$ at the center of the interface
(where $p_i=p_j=1/2$).

\subsection{Relation to sharp-interface models}

The model being completely specified, let us now relate 
its parameters to the quantities that usually appear in 
sharp-interface theories. To this end, it is useful to give 
a few more details on the properties of this phase field 
model; for a more exhaustive discussion, see Ref.~\cite{Folch05}. 

The free energy functional is constructed such that $p_k = 0$ 
is a stable solution along each $i-j$  interface both at 
equilibrium and out of equilibrium, which means
that each two-phase interface can be described by a single 
phase-field variable: since $p_k=0$, $p_i$ or $p_j$ can be
eliminated using the constraint $p_\alpha+p_\mother+p_\beta=1$.
Furthermore, the special form chosen for the coupling between
the concentration and phase fields yields a particularly
simple expression for the chemical potential, 
\be
\mu = \frac{\delta{\cal F}}{\delta c} = 
\tilde\lambda \left[c-g({\bf p})\right].
\ee
For an equilibrium interface, we have $\partial_t p_i = 0$
for all phase fields, and the chemical potential is constant.
Let us first examine a grain boundary, that is, an interface 
between the $\alpha$ and $\mother$ phases.
Since according to the above properties, $p_\beta=0$ along
the whole interface, we have $g({\bf p})\equiv 0$ and
hence $\mu$ constant implies $c$ constant: there is no
grain boundary segregation. 

Next, consider a planar interphase boundary between phases 
$\mother$ and $\beta$. Since we have $p_\alpha\equiv 0$,
$\mu$ can be expressed as a function of $c$ and one of
the phase fields, say $p_\beta$. This yields
\be
\mu = \tilde\lambda \left[c - g_\beta(p_\beta)\right]
\label{muequation}
\ee
with $g_\beta(p)=p^3(10-15p+6p^2)$. The equation for
the phase field $p_\beta$ for a planar interface normal 
to the $x$ direction becomes
\be
0 = \partial_t p_\beta = W^2 \partial_{xx} p_\beta 
  - f_\beta'(p_\beta) - \mu g'_\beta(p_\beta)
\label{pequation}
\ee
with $f_\beta(p)=2p^2(1-p)^2\left[1+a_\alpha p(1-p)\right]$;
the equivalent equation for the $\alpha$-$\beta$ interface
can be obtained by omitting the term proportional to $a_\alpha$.
Since the chemical potential is equal to zero at two-phase 
equilibrium, all terms depending on $c$ disappear from 
Eq.~(\ref{pequation}), which hence becomes an equation for the 
phase field only. Our phase-field model has been specifically
designed to achieve this exact decoupling, which is not a
general property of multi-phase-field models \cite{Nestler05}.
For $a_\alpha=0$, the solution of Eq.~(\ref{pequation}) is the
standard hyperbolic tangent profile; for $a_\alpha\neq 0$, a 
modified equilibrium front profile $p_\beta^0(x)$ is obtained. 
In both cases, the equilibrium concentration profile can
then be obtained from Eq.~(\ref{muequation}) as 
$c(x)= g_\beta(p_\beta^0(x))$.

Several consequences arise from the structure of the
model: first, the surface tensions of the interfaces
can be calculated from the phase-field part of the
free energy alone; therefore, the surface tensions
are independent of the concentration. They can be
calculated by standard procedures in the form of an integral,
\be
\sigma_{\mother\beta} = 
2\sqrt{2}WH_p \int_{0}^{1} p(1-p)\sqrt{1 + a_\alpha p(1-p)} dp,
\label{sigma}
\ee
where $\sigma_{\alpha\beta}$ and the grain boundary
energy $\sigma_{gb}$ are obtained by setting $a_\alpha=0$,
which yields $\sigma_{\alpha\beta}=\sigma_{gb}=WH_p\sqrt{2}/3$.
In the present study, we restrict ourselves to the case where 
these two surface energies are equal; however, the general case
can be easily treated by adding another term to the
free energy functional (see Ref.~\cite{Folch05}).
Furthermore, standard calculations yield the Gibbs-Thomson
relation for the $\alpha$-$\beta$ and $\mother$-$\beta$ interfaces,
\be
\mu_{\rm int} = d_{\alpha\beta}\kappa,
\label{GiTho}
\ee
\be
\mu_{\rm int} = d_{\mother\beta}\kappa,
\ee
where $\kappa$ is the interface curvature (counted positive
when the $\beta$ domain is convex) and the capillary lengths are
given by
\be
d_{i\beta}=\frac{\sigma_{i\beta}}{H_c}=\frac{\sigma_{i\beta}}{\partial f^2/\partial c^2}\quad(i=\alpha,\mother).
\ee
Since there is no grain boundary segregation (and hence no 
solute drag effect), grain boundaries exhibit the standard
motion by curvature,
\be
V_n = - \sigma_{gb} M_{gb} \kappa
\label{eq:mbycurvature}
\ee
where $V_n$ is the normal velocity of the grain boundary, and
the grain boundary mobility is given by
\be
M_{gb} = \frac{W}{\tau H_p}.
\label{Mgb}
\ee
Note that the product $\sigma_{gb}M_{gb}$ has the dimension of
a diffusion coefficient and scales as $W^2/\tau$; this is
actually the diffusion coefficient that appears in the equations
of motion for the phase fields.

\begin{figure}
\centerline{\epsfig{figure=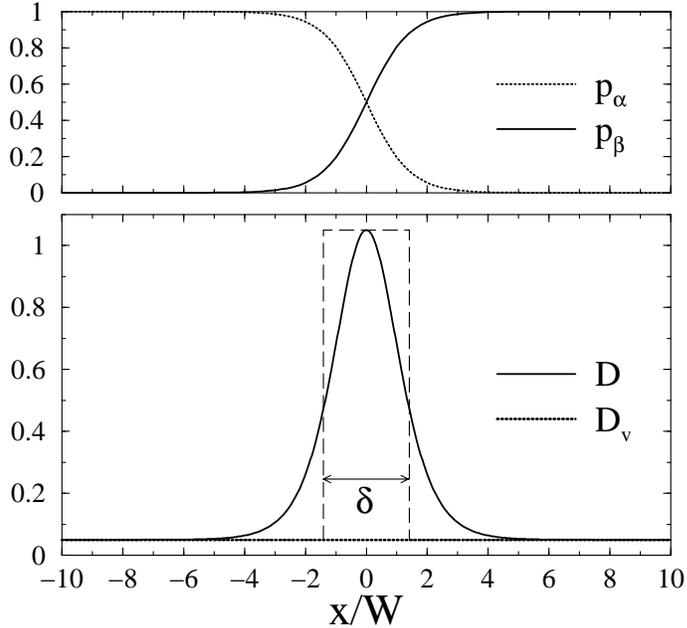, width=9cm}}
\caption{\label{fig:diff}Relation between the diffusivity function
of the phase-field model and the conventional sharp-interface picture.
See text for details.}
\end{figure}

Finally, let us comment on the surface diffusion coefficients.
In the standard picture of grain boundary diffusion,
a grain boundary or interface is seen as a region of 
well-defined width $\delta$ in which the diffusivity
markedly differs from the bulk value; the diffusivity
hence formally has a jump at the sharp boundary of
the interface zone. In contrast, in the phase-field
picture the transition between bulk and ``surface''
value is smooth. To make contact between the two pictures,
it is useful to proceed via a Gibbs construction, as
illustrated in Fig.~\ref{fig:diff}. In the upper panel, the
profiles of the phase fields in an $\alpha$-$\beta$
interface are shown. In the lower panel, the diffusivity 
function for $\dbulk=0.05$ and $\dsurfab=1$ is displayed
together with a step function that has a certain width
$\delta$ which is defined by the relation
\be
\dsurfab\delta = \int_{-\infty}^\infty dx\; D[{\bf p}(x)]-\dbulk ,
\label{Dgibbs}
\ee
that is, the step function and the smooth diffusivity
function represent the same total {\em excess diffusivity}
with respect to the bulk value. For the standard hyperbolic
tangent ($a_\alpha=0$), we obtain analytically $\delta = 2\sqrt{2}W$;
for $a_\alpha\neq 0$, the value of $\delta$ has to be obtained
numerically.

Let us briefly comment on how the model parameters can be
determined to simulate a given alloy system. The constant
$H_c$ is fixed by thermodynamics, since it depends only on
the free energy curve and the equilibrium compositions. The
capillary lengths can be obtained from this constant and the
surface tensions. The latter also fixes the product $WH_p$
through Eq.~(\ref{sigma}). The interface thickness $W$ can either
be fixed using structural information, or treated as a free
parameter; in both cases, once a value for $W$ is fixed, the
parameters $H_p$, $K=W^2H_p$ and $\tilde\lambda=H_c/H_p$ are 
fixed. Regarding the surface diffusivities, usually only their
product with $\delta$ is known. However, once $W$ chosen in
the model, $\delta$ can be calculated from Eq.~(\ref{Dgibbs}),
and thus the value of the surface diffusivities can be fixed.
Finally, $\tau$ can be determined through Eq.~(\ref{Mgb})
from the value of the grain boundary mobility.

It is convenient to non-dimensionalize the equations. We choose
as units of length, time, and free energy density $W$, $\tau$,
and $H_p$. In the final model equations, the only remaining 
parameters are the constant $a_\alpha$ in $f_{TW}$ which influences
the surface tension of the $\mother$-$\beta$ interface, the
constant $\tilde\lambda$, and the dimensionless solute diffusion
coefficients; for example, the scaled grain boundary diffusion
coefficient $\tilde\dsurfma$ reads
\be
\tilde\dsurfma = \frac{\dsurfma\tau}{W^2}= \frac{\dsurfma}{W H_pM_{gb}}.
\label{eq:Dnondim}
\ee
Note that this dimensionless combination can be related to the
dimensionless parameter $\beta$ of Cahn's theory \cite{Cahn59}.
For simplicity, we will drop the tildes for the remainder of 
the paper.

\section{Simulation setup and parameters}
\label{sec:simulation}

\begin{figure}
\centerline{\epsfig{figure=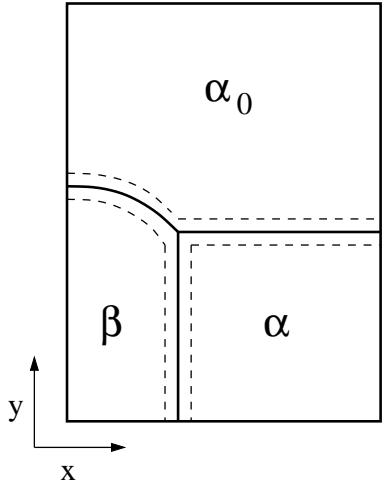, width=5cm}}
\caption{\label{fig:simbox}Sketch of the geometry of the simulation 
box. Half of a precipitate is simulated, with reflection boundary 
conditions on all sides, except ahead on the growth front, where the
concentration and the phase fields are kept fixed to the values
corresponding to the supersaturated $\alpha_0$ phase. The lateral
box size is $L/2$, where $L$ is the interlamellar spacing. The
dashed lines indicate the limits of the diffuse interfaces. The
drawing is not to scale: in most of the simulations, the interfaces
are thinner.
}
\end{figure}

The equations for the phase fields and the concentration are 
discretized using finite-difference formulas, and integrated in 
time using an explicit Euler algorithm. Since we are interested
in this study in strictly periodic lamellar arrays only, we can 
take advantage of the planes of symmetry which are present 
in the center of each lamella, and compute only half of a 
lamella pair, as sketched in Fig.~\ref{fig:simbox}, with 
reflection boundary conditions at the lateral sides. The lamellar
spacing $L$ is hence fixed by the size of the simulation box.

We start our simulations from a flat grain boundary in
contact either with a round precipitate of $\beta$ phase, or with 
a pre-existing $\beta$ lamella. The values of the concentrations
are initially set to the equilibrium values in the $\alpha$
and $\beta$ phase ($c=0$ and $1$, respectively), and to the 
chosen supersaturation $\Delta$ in the $\mother$ 
phase ($c=\Delta$). In order to speed up the simulations,
the box is relatively small in the growth direction and
is moved periodically to maintain the growth front in
its center. The box size is always large enough to obtain 
results that are independent of the box size. The growth velocity
of the precipitate and the grain boundary are monitored
as a function of time. Once a steady state is obtained, 
it can be used as an initial condition for subsequent 
runs with different parameters. This considerably speeds 
up the convergence to steady-state solutions.

Since we are interested in generic features of discontinuous
precipitation, we make some reasonable choices for the parameters
rather than to attempt to model a particular alloy system. We
set $\tilde\lambda=1$, which yields a ratio of capillary lengths
and interface thickness of order unity. For the choice of
surface tensions and surface diffusivities, we take into
account some findings of Ref.~\cite{Brener99} which helps
to narrow down the field of investigation. First, the contact
angles at the trijunction point have to be such that the $\beta$
precipitate is convex along the entire $\alpha$-$\beta$
interface. For a steady state with a flat grain boundary such 
as depicted in Fig.~\ref{fig:simbox}, this is equivalent to the 
requirement that the angle between the grain boundary
and the $\alpha$-$\beta$ interface has to be smaller than
90$^\circ$. The physical foundation of this condition
is relatively easy to understand: if the $\alpha$-$\beta$
interface develops overhangs, in a steady-state solution
this implies that the overhanging parts of $\beta$
have to dissolve behind the front. However, since 
at least some parts of this interface have to be
concave, they have a lower chemical potential than
the surrounding flat or convex parts of the interface,
which implies that $\beta$ should grow rather than
dissolve. This is indeed what we observe in simulations 
where the aforementioned condition is not satisfied:
the $\beta$ precipitate grows sideways and slows down;
no steady-state growth is reached.

The conditions on the surface tensions can be obtained from
Young's law at the trijunction point. In our model,
$\sigma_{\alpha\beta}=\sigma_{gb}$, and hence we must have
$\sigma_{\mother\beta}/\sigma_{gb}>2\cos(\pi/4) = \sqrt{2}$.
We choose $a_\alpha=9$, which yields 
$\sigma_{\mother\beta}= 0.7856 \, H_pW = 1.666 \,\sigma_{gb}$.
A consequence of this choice which has some practical
implications is that the $\mother$-$\beta$ interface is
thinner than the others. This forces us to use a rather
fine discretization of $\Delta x=0.4 \, W$. Even with this
value, some grid effects remain visible, but a further
refinement does not appreciably change the simulation
results.

The parameters we focus on in this investigation are the
diffusivities $\dbulk$, $\dsurfma$, $\dsurfmb$, and $\dsurfab$. 
Of the surface diffusivities, the first two ones 
control the flux of solute along the growth
front, whereas the latter controls the diffusion in
the interface {\em behind} the front. In Ref.~\cite{Brener99},
it was found that the value of $\dsurfab$ has a strong
influence on the front velocity, and steady-state solutions
could be found only below a critical value for $\dsurfab$.
Therefore, we decided to always set $\dsurfma=\dsurfmb$,
but to keep $\dsurfab$ as an independent parameter.
There are thus three relevant independent diffusion
coefficients that need to be investigated: $\dbulk$,
$\dsurfma$, and $\dsurfab$.

\section{Results}
\label{sec:results}

\subsection{General remarks}

\begin{figure}
\centerline{\epsfig{figure=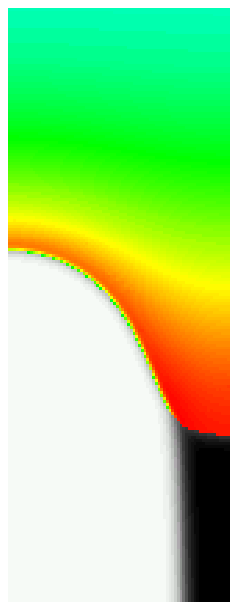,width=2.5cm}\hspace{2.5mm}
            \epsfig{figure=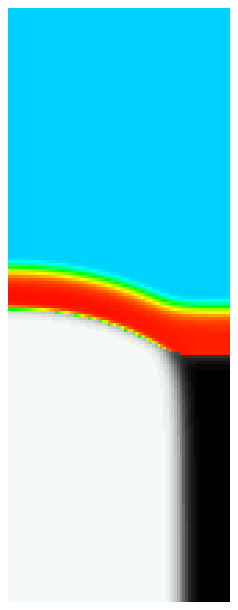,width=2.5cm}\hspace{2.5mm}
            \epsfig{figure=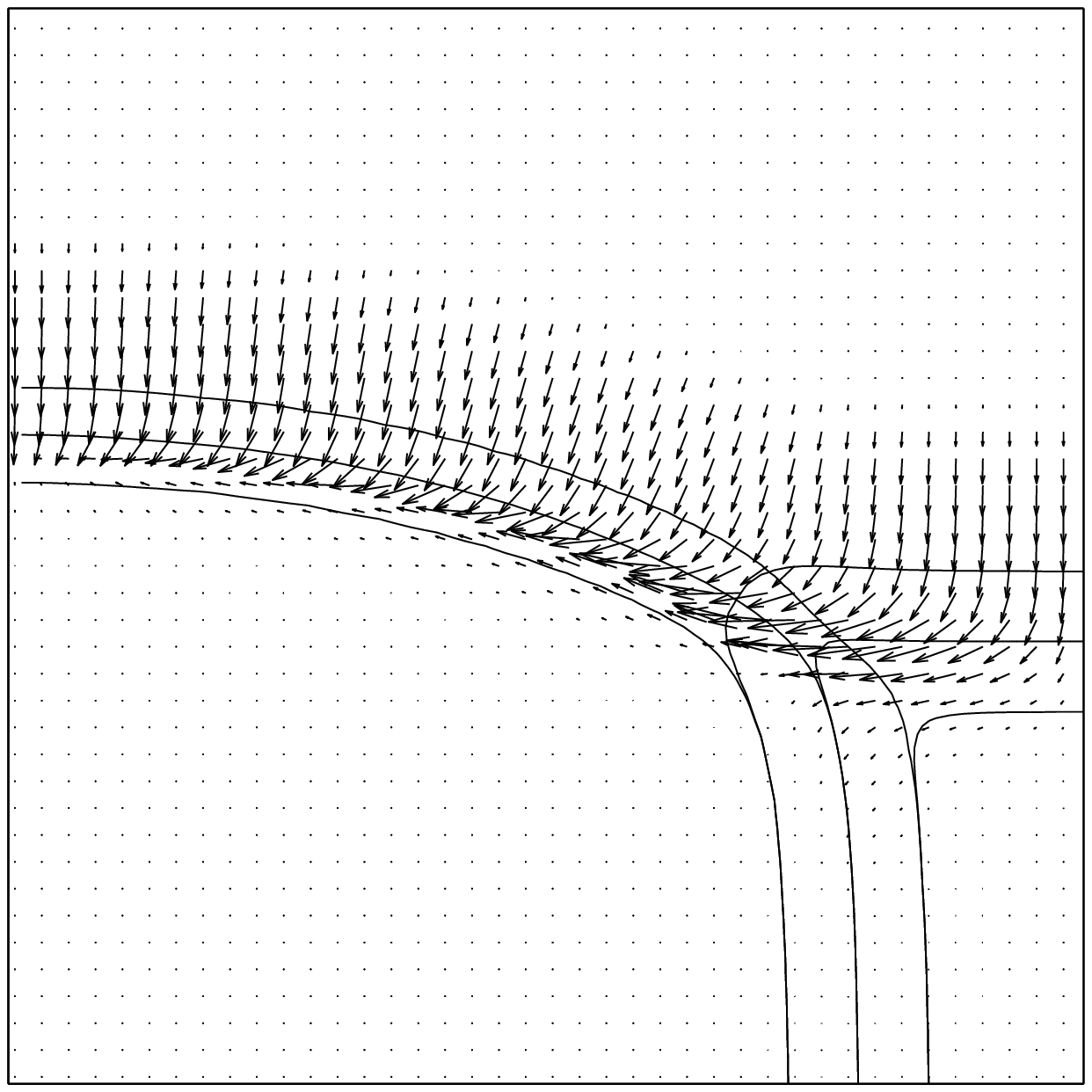,width=6.5cm}}
\caption{\label{fig:global}Left: snapshot of a simulation with
$\dbulk=1$ and $\dsurfma=\dsurfmb=\dsurfab=0$. Middle: 
$\dbulk=10^{-6}$, $\dsurfma=\dsurfmb=1$ and $\dsurfab=6\times 10^{-3}$.
We recall that all diffusion coefficients are scaled according
to Eq.~(\protect\ref{eq:Dnondim}).
Right: blowup of the front region; the arrows represent the diffusion
currents. In all cases, $\Delta=0.8$ and $L/W = 64$.}
\end{figure}

In our simulations, we have identified two distinct regimes: 
growth limited by volume diffusion and by surface diffusion. 
To illustrate the difference, we show in Fig.~\ref{fig:global} 
two representative snapshots of steady-state precipitates
corresponding to the two regimes. In the bulk-diffusion 
limited case, the diffusion field extends far into the bulk
of the mother phase, and the precipitate is pointy, that is, 
the curvature is greatest at the precipitate tip. In contrast, 
in the surface-diffusion limited case, the precipitate is much 
flatter, and the diffusion field is localized in the vicinity of 
the interfaces. The latter point is further illustrated by the plot
to the right, which shows a map of the diffusion currents.
Globally, the growth is much faster in the bulk-diffusion
limited case.

\subsection{Bulk-diffusion limited growth}

Since the rate-limiting step in discontinuous precipitation
is surface diffusion, we will present here only our most
important findings about the bulk-diffusion-limited regime;
more details will be given elsewhere.

Let us first consider the purely bulk-diffusion-limited case,
that is, $\dbulk=1$ and $\dsurfma=\dsurfmb=\dsurfab=0$. Note
that with our definition of the surface diffusivity, zero
surface diffusivity simply means that the diffusivity in the
interface region is the same as in the bulk. In this limit,
the problem is closely related to the growth of a crystalline
finger in a channel, which has been considered in numerous
studies of solidification \cite{Brener88,Brener93channel}. 
Indeed, if the precipitates grow from the mother phase without 
the presence of the grain boundary and hence of the second grain, 
the two problems are completely equivalent. For crystal
growth in a channel, it is known \cite{Brener88,Brener93channel}
that steady states can only 
exist for a channel width exceeding a critical value which 
depends on the supersaturation. At this critical width,
the branch of stable steady-state solutions exhibits a
fold singularity: it merges with a second branch of unstable 
solutions. For a channel width (which is equivalent to the 
spacing here) above this value, the growth velocity first 
increases with increasing width, goes through a maximum, 
and then decreases until the steady-state fingers become 
unstable against tip-splitting.

We observe qualitatively the same behavior, but the values 
of the growth velocity also depend on the properties of the
grain boundary. In particular, the grain boundary mobility
plays an important role. In the snapshot picture of Fig.~\ref{fig:global},
it can be seen that the grain boundary is slightly curved.
If the grain boundary mobility is changed at fixed growth velocity, 
according to Eq.~(\ref{eq:mbycurvature}) the curvature of the grain 
boundary and hence the contact angles at the trijunction are
modified; this, in turn, modifies the shape of the precipitate 
tip and the surrounding diffusion field. When the grain boundary 
becomes more sluggish, $\dbulk>1$ (we recall 
that the scaled diffusion coefficient is proportional to
the ratio of the solute diffusivity and the grain boundary
mobility), it falls even further behind the precipitate tip
than shown in the snapshot of Fig.~\ref{fig:global}, 
and the shape of the precipitate tip approaches the one
of a crystalline finger in a channel. In contrast, when the 
grain boundary mobility is increased ($\dbulk<1$), the
curvature of the grain boundary decreases and the precipitate 
becomes flatter, which leads to a lower growth velocity.
Below a certain critical value of $\dbulk$ that depends
on the spacing $L$, no steady-state growth is possible.
If, in addition, surface diffusion is included for otherwise
unchanged parameters, the growth velocity always increases, 
but the qualitative behavior described above remains unchanged.

\begin{figure}
\centerline{\epsfig{figure=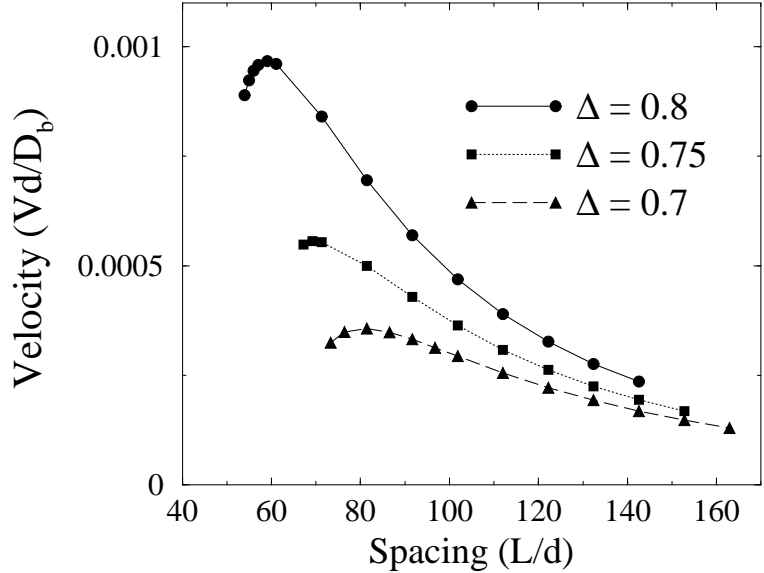, width=10cm}}
\caption{\label{fig:VvsL}Steady-state velocity versus lamellar
spacing for different supersaturations. $\dsurfma=\dsurfmb=1$,
$\dsurfab=10^{-3}$, $\dbulk=10^{-6}$. Lengths are scaled by the
capillary length $d\equiv d_{\mother\beta}=0.7856\, W$, and 
times by $d^2/\dsurfma$.}
\end{figure}

\subsection{Surface-diffusion limited growth}

The more relevant case for the description of discontinuous
precipitation is growth limited by surface diffusion, which
occurs when $\dbulk\ll\dsurfma$. In addition, as will be shown
in more detail below, we must have $\dsurfab\ll\dsurfma$.
Globally, the growth velocities are much slower than in the 
bulk-diffusion limited case, and therefore the grain boundary 
mobility has no noticeable influence in this regime (the 
curvature of the grain boundary always remains small).

\begin{figure}
\centerline{\epsfig{figure=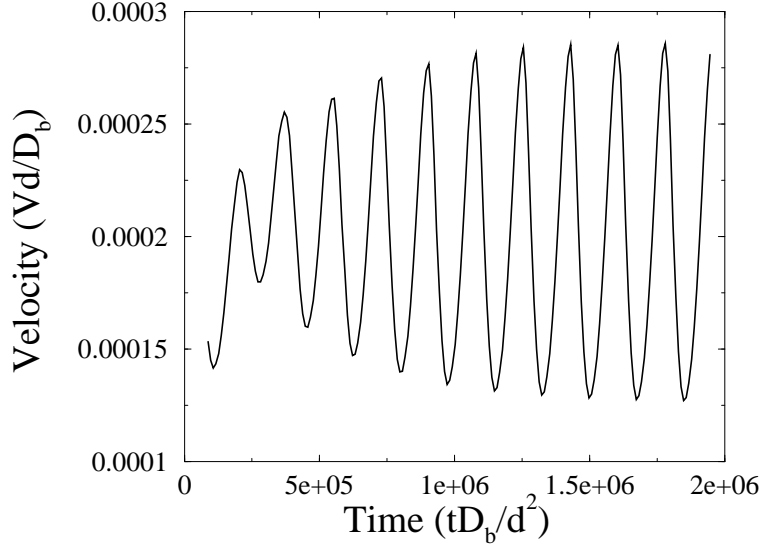, width=10cm}}
\caption{\label{fig:lcycle}Velocity of the grain boundary
as a function of time for $\Delta=0.8$ and $L/d=152.75$; 
$\dsurfma=\dsurfmb=1$, $\dsurfab=10^{-3}$, $\dbulk=10^{-6}$.
The velocity starts to oscillate, and the amplitude of the oscillation
saturates after some time: the system has reached a limit cycle.}
\end{figure}

For fixed diffusion coefficients, the precipitate growth 
velocity $V$ depends on the spacing $L$ and the supersaturation 
$\Delta$. In Fig.~\ref{fig:VvsL}, we plot the growth velocity 
versus spacing for different supersaturations. As for
bulk-diffusion limited growth, the velocity-versus-spacing 
curve has a maximum for a certain spacing. For low spacing,
the curve ends with a diverging slope at a finite value of
the growth velocity. This indicates that the lower limit
for steady-state spacing corresponds to a fold singularity,
as predicted in Ref.~\cite{Brener99}. For spacings below
this critical value, no steady-state solution can be found 
any more. Instead, the growth front velocity decreases with
time and the precipitate grows in the lateral direction. 
For spacings larger than the maximum velocity spacing, 
$V$ decreases with increasing spacing until an instability 
is reached: all of the precipitate velocity, the precipitate 
width and the velocity of the grain boundary 
start to oscillate until the dynamics reaches an
oscillatory limit cycle, as illustrated in Fig.~\ref{fig:lcycle}.

For a fixed spacing, the velocity increases monotonously
with the supersaturation, but no simple scaling law was found. 
Note that, for fixed spacing, steady-state growth is possible 
only in a range of supersaturations, the minimum and maximum values 
of which are set by the fold singularity and the onset of the 
oscillatory instability, respectively.
All characteristic spacings (the minimum spacing, the maximum 
velocity spacing, and the spacing for the onset of the oscillations)
increase with decreasing supersaturation. Since simulations
are quite time-consuming due to the slow growth velocities, we 
have not investigated even lower supersaturations, which would 
be necessary to determine the scaling of the characteristic 
spacings with supersaturation.

\begin{figure}
\centerline{\epsfig{figure=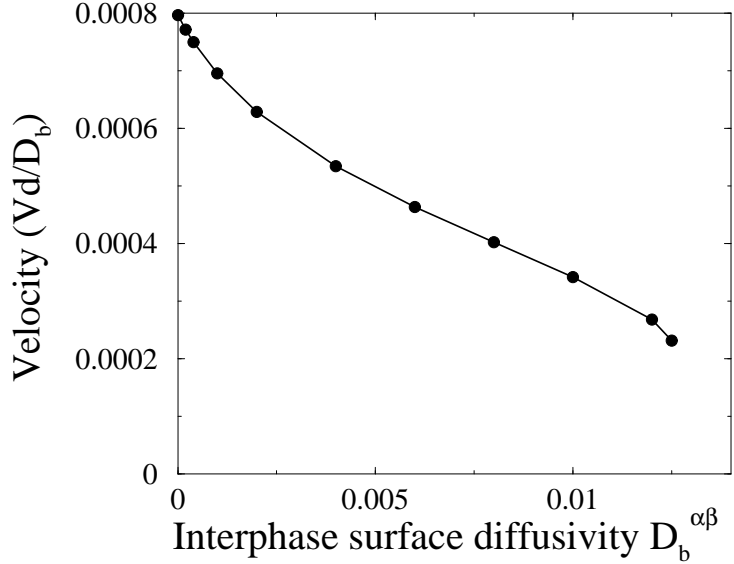, width=10cm}}
\caption{\label{fig:VvsDs}Steady-state velocity as a function
of the surface diffusivity of the $\alpha$-$\beta$ interface,
$\dsurfab$. The other parameters are: $\Delta=0.8$, $L/d=81.5$,
$\dsurfma=\dsurfmb=1$,$\dbulk=10^{-6}$.}
\end{figure}

Next, we investigate the influence of the surface diffusivity in the
interphase boundary {\em behind} the growth front on the precipitate
growth velocity: $\dsurfab$ is varied while all the
other parameters and the spacing are kept constant.
In agreement with the predictions of Ref.~\cite{Brener99}, 
Fig.~\ref{fig:VvsDs} reveals that the growth velocity decreases 
with increasing $\dsurfab$; above a certain critical value, no 
steady-state solutions exist any more. In contrast, no new behavior
appears when the value of $\dsurfab$ is lowered; it can
even be set to zero.

\begin{figure}
\centerline{\epsfig{figure=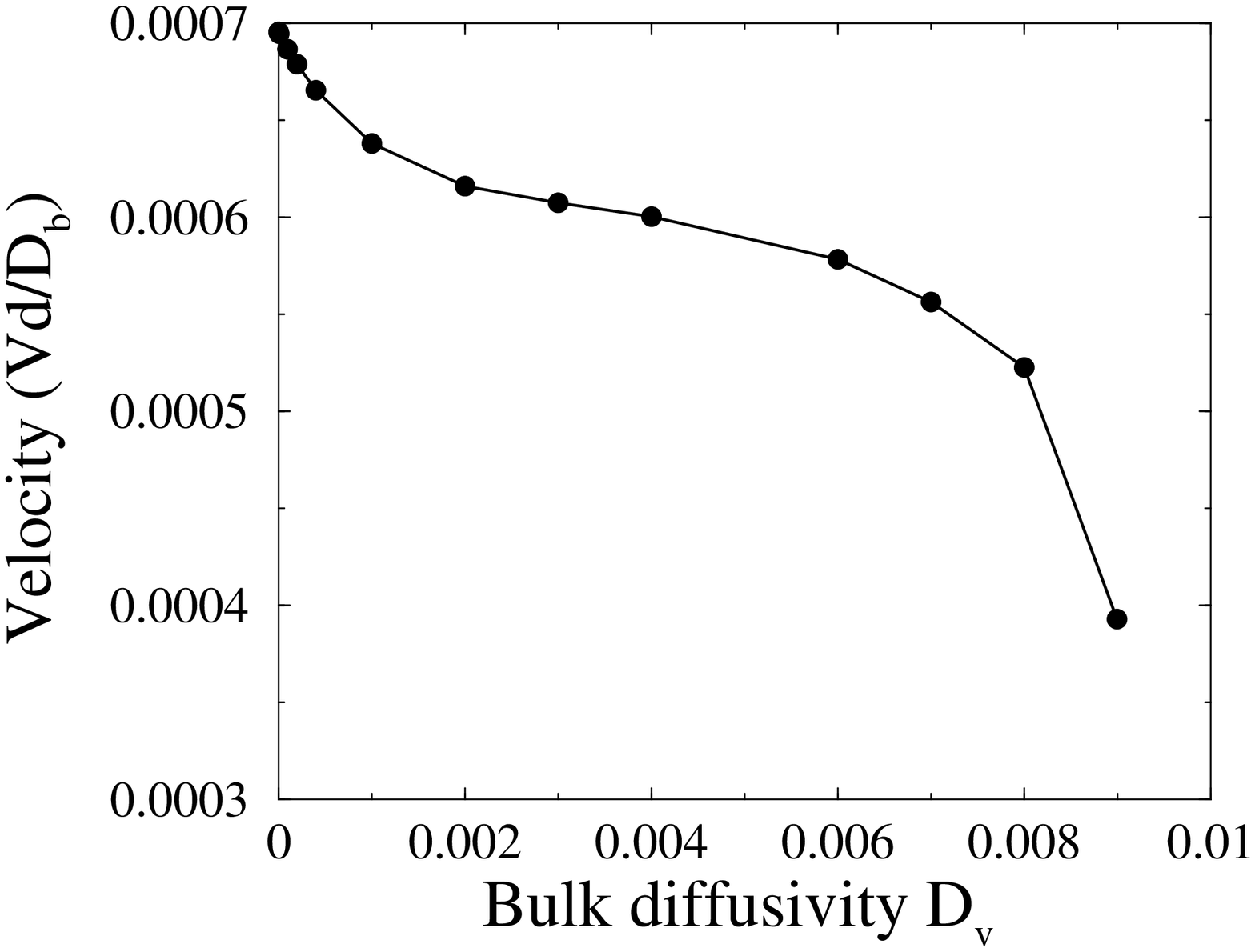, width=10cm}}
\caption{\label{fig:VvsDb}Steady-state velocity as a function
of the bulk diffusivity. The other parameters are: $\Delta=0.8$, 
$L/d=81.5$, $\dsurfma=\dsurfmb=1$,$\dsurfab=10^{-3}$.}
\end{figure}

A surprising behavior is observed when $\dbulk$ is increased, 
as shown in Fig.~\ref{fig:VvsDb}. Contrary to what one
might expect, an increase of the bulk diffusivity {\em slows
down} growth. Above a critical value of $\dbulk$, 
no steady-state solution exists any more. This is
especially noteworthy because it implies that
there is no continuous branch of solutions which
links the surface-diffusion-limited and the
bulk-diffusion-limited regimes.

\subsection{Comparison to sharp-interface models}

In order to compare our simulation results to the predictions
of the sharp-interface theory, we have solved numerically the 
complete system of equations developed in Ref.~\cite{Brener99} that 
implicitly gives the velocity as a function of spacing; the values
of the supersaturation attainable in our simulations are too high for 
the explicit closed-form approximations given in Ref.~\cite{Brener99}
to be applicable. The comparison of the predicted to the simulated 
growth velocities reveals that some features of our results are qualitatively 
well predicted by the theory: (i) the existence of a fold singularity that 
sets a lower limit for the spacing, (ii) the initial increase of velocity 
with increasing spacing, and (iii) the strong influence of the surface 
diffusivity in the interphase boundary behind the front on the growth
velocity. However, there are also some strong discrepancies: the 
occurrence of a maximum velocity at a well-defined spacing and the
subsequent decrease of the velocity with increasing spacing, 
as well as the occurrence of the oscillatory 
instability, are not captured by the theory of Ref.~\cite{Brener99}. 
Moreover, there are important differences in the magnitudes of the
velocities and spacings: the theory predicts growth velocities that 
are about 40 to 50 times larger than the ones observed in our 
simulations, whereas the minimum steady-state spacing found in 
our simulations is about three times larger than the predicted one.

\begin{figure}
\centerline{\epsfig{figure=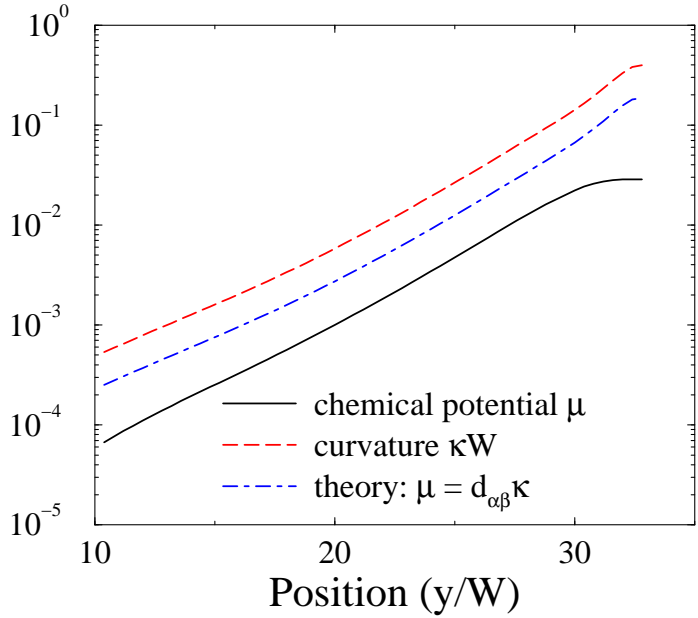, width=9cm}}
\caption{\label{fig:theocomp}Curvature of the $\alpha$-$\beta$
interface and chemical potential at the interface as a function 
of the vertical coordinate $y$. The chemical potential obtained
from the local equilibrium assumption is also shown.}
\end{figure}

It is interesting to investigate what is the reason for these 
discrepancies. To this end, we choose to analyze one particular
simulation and to check which ingredients of the
sharp-interface theory are a good description of our
simulations, and which have to be revised.
We focus on the $\alpha$-$\beta$ interface,
since the solution of the complete free boundary
problem in Ref.~\cite{Brener99} provides a particularly
simple prediction for its shape: the curvature of the 
interface decreases exponentially with the distance from the 
trijunction point, $\kappa\propto \exp[q(y-y_t)]$, where $y_t$ 
is the $y$ coordinate of the trijunction point, and $q$
is the inverse of a decay length, which can be related
to the model parameters \cite{Brener99}. In Fig.~\ref{fig:theocomp} 
we plot the interface curvature as a function of $y$ and
find indeed an exponential decay.

The sharp-interface model assumes local equilibrium at the 
$\alpha$-$\beta$ interface according to Eq.~(\ref{GiTho}). 
The solute diffusion along the interface is then driven by 
the curvature gradient, which creates a chemical potential
gradient. To check these assumptions,
we extract the chemical potential at the center 
of the interface and find indeed an exponential decay,
as expected. However, the curvature and the chemical
potential extracted from our simulations are {\em not}
in agreement with the Gibbs-Thomson law of local
equilibrium, Eq.~(\ref{GiTho}).

\begin{figure}
\centerline{\epsfig{figure=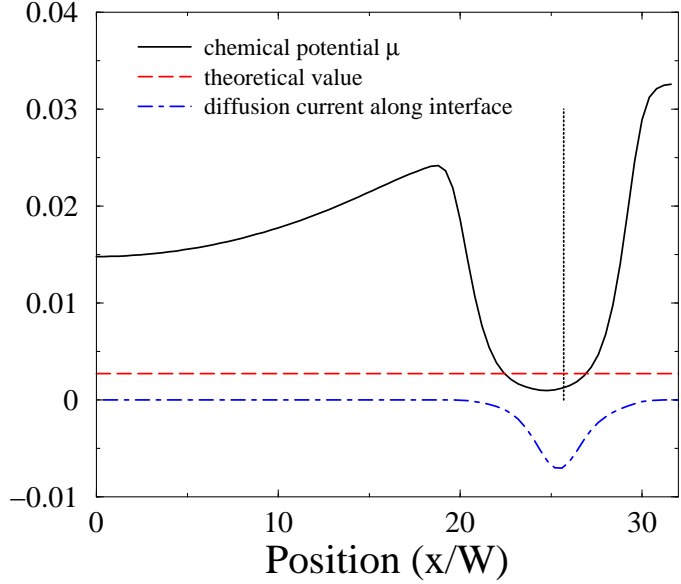, width=9cm}}
\caption{\label{fig:abinter}Plot of various quantities as a function
of $x$ for a $y$ coordinate corresponding to a position halfway
between the trijunction point and the lower system boundary.
See text for details.}
\end{figure}

To get a more detailed picture, we plot in 
Fig.~\ref{fig:abinter} a profile of various quantities
along a horizontal line that cuts through the $\alpha$-$\beta$
interface at some distance behind the trijunction point.
The current density in the $y$ direction is shown as a 
dash-dotted line. Since the bulk diffusion is very slow, this 
current is close to zero in both the $\alpha$ and $\beta$ phases.
It exhibits a peak in the diffuse interface, with a
negative sign: this is
the surface current which is driven by the curvature
gradient along the $\alpha$-$\beta$ interface from
the trijunction region to the region far behind the front
where the interface is flat. The shape of the peak is
as expected from the diffusivity function shown in
Fig.~\ref{fig:diff}, which indicates that the general
picture of a smooth ``surface current density'' is correct.

The chemical potential, shown as a full line, has smooth 
variations in both bulk phases and is positive.
This profile is essentially set by the growth history:
since bulk diffusion is slow, the concentration (and
hence the chemical potential) remains approximately at
the value it had immediately after the passage of the 
growth front. In contrast, $\mu$ strongly varies in the interface
region and exhibits a ``dip'': in the center of the interface,
it is below the value predicted by the Gibbs-Thomson relation,
whereas in the regions adjacent to the bulk
it is above. Thus, not only the chemical potential
in the interface does not satisfy the Gibbs-Thomson relation,
but even the concept of a uniform ``interface chemical
potential'' cannot be maintained. 

Qualitatively, this ``dip'' in the chemical potential is 
due to the fact that solute can be rapidly transported 
along the interface, and hence the chemical potential
can change much faster in the center of the interface than
in the bulk. Since the chemical potential of a flat interface
is zero, and hence lower than any value occurring in the
bulk, the diffusion along the interface ``drains'' solute
from the surrounding bulk.
Note that, as a consequence, the local values of the
chemical potential are history-dependent, even inside
the diffuse interface where the diffusion is fast.
This can be recognized in Fig.~\ref{fig:abinter}, in which 
the position of the center of the interface is marked by 
a vertical dotted line. Clearly, the 
``dip'' in the chemical potential is asymmetric. This can 
be explained by the fact that this interface has moved
slightly ``to the right'' (toward increasing $x$) since 
the passage of the trijunction point. Therefore, the points 
on the left side of the interface have already been drained
by the surface current, whereas the points to the right have not.

\section{Discussion}
\label{sec:discussion}

\subsection{Influence of the diffusivities}

One of the surprising results of the present study is that
the two regimes where growth is limited by bulk and surface 
diffusion, respectively, seem to be distinct in the sense 
that we have found no continuous family of solutions 
connecting them.

A possible explanation comes from the geometric constraints.
Let us first examine growth limited by surface diffusion and
suppose furthermore that there is no diffusion in the 
$\alpha$-$\beta$ interface behind the growth front. Then, the growth
of the $\beta$ precipitate requires that solute is transported
from the $\alpha$ to the $\beta$ lamella along the growth front; 
a current of solute atoms must hence flow from 
the center of the grain boundary
to the center of the $\beta$ lamella (from right to left in
Fig.~\ref{fig:simbox}). Even though the local equilibrium
assumption is not valid, Fig.~\ref{fig:theocomp} shows that
the local chemical potential still increases with curvature.
Hence, such a current can flow only if the curvature
decreases from the sides to the center of the $\beta$ lamellae;
as a consequence, the curvature must exhibit a minimum in
the center of the $\beta$ lamella.

The decrease of the growth velocity with the increase of the 
diffusivity in the $\alpha$-$\beta$ interface is easily
understood: when diffusion along this interface takes place,
a part of the solute atoms is drained toward the flat
parts of the interface far behind the growth front where
the chemical potential is lowest; this material is lost
for the forward growth of the precipitate, and the surface 
current to the center of the $\beta$ lamella decreases.

When bulk diffusion is allowed, there is an alternative diffusion
path through the volume, which on first thought should accelerate
growth. However, diffusion in the bulk also leads to morphological
instability: the concentration gradients are enhanced around
protruding parts of the interface, which hence advance faster
than flat parts, leading eventually (in the regime limited by
volume diffusion) to the emergence of cellular precipitates
which exhibit a maximum of curvature at the tip. A possible
explanation of the decrease of the growth velocity with increasing
bulk diffusivity is that the bulk diffusion leads to an increase
of curvature at the precipitate tip, which decreases the driving
force for lateral surface diffusion. The two diffusion mechanisms
hence play antagonistic roles. We found steady states only when
one of these mechanisms is strongly dominant over the other. 
However, we have only explored a small part of the parameter
space spanned by $L$, $\Delta$, $\dbulk$, $\dsurfma$, and
$\dsurfab$, and hence we cannot exclude that there is a path
which connects the two types of solutions. Further studies are
needed to clarify this point.

\subsection{Breakdown of local equilibrium}

To understand the breakdown of local equilibrium, let us revisit
some of the fundamental ideas behind the local equilibrium
assumption. The concept of ``local equilibrium'' implies that
there is a separation of length and time scales: in a small part
of the system, some fast processes can establish and maintain a
local thermodynamic equilibrium, whereas the entire system evolves
on large length scales and slow time scales. This provides
``adiabatic'' changes in the boundary conditions on the small
scale. In the case considered here, namely slow bulk diffusion
and fast surface diffusion, these definitions become ambiguous.
Indeed, since the diffusivity rapidly falls to a small value
in the bulk, it takes much longer for a solute atom to diffuse
from one side of an interface to the other than to diffuse
along the interface by a distance which is much larger than
the interface thickness.

Therefore, the ``small volume element'' to be considered for
local equilibrium is strongly anisotropic. To be more precise,
consider the diffusion times associated with diffusion across 
and along the interfaces, $t_\perp$ and $t_\parallel$,
\be
t_\perp = \frac{W^2}{\dbulk}
\label{eq:dperp}
\ee
\be
t_\parallel = \frac{l^2}{\dsurfma}.
\ee
We have used the value of the bulk diffusivity in the first
expression because the rate-limiting step for diffusion across
the interface is the crossing of the outer regions of the
interface where the diffusivity is the lowest and almost equal
to the bulk diffusivity. In the second expression, $l$ is an
as of yet unknown length scale. Equating the two time scales,
we obtain
\be
l=W\sqrt{\frac{\dsurfma}{\dbulk}};
\ee
obviously, $l$ can be much larger than $W$.

The concept of local equilibrium remains valid only if the
``external conditions'' -- here, the curvature and interface
velocity -- remain almost constant over this distance. The 
characteristic length scale for variations of curvature and 
interface velocity is the lamellar spacing $L$; hence we obtain 
the condition
\be
W\sqrt{\frac{\dsurfma}{\dbulk}} \ll L
\ee
for the validity of local equilibrium.

In our simulations, $\sqrt{\dsurfma/\dbulk}=10^3$, whereas typical
system sizes are $L\sim 50\,W$; therefore, this condition is
not satisfied. To check whether this corresponds to a real
experimental situation, we consider the alloy Al-Zn which has
been extensively studied \cite{Yang88}. In the temperature range 
between 400 and 500 K, spacings are of the order 100 nm, and growth
velocities of the order $10^{-7}$ m/s. The bulk diffusion coefficient
varies from $\sim 10^{-16}$ to $\sim 10^{-19}$ m$^2$/s, whereas
the triple product $s\delta\dsurfma$, where $s$ is the segregation
factor, is of order $10^{-20}-10^{-21}$ m$^3$/s. Assuming
$s\approx 1$ and $\delta = 1$ nm, the ratio $\dsurfma/\dbulk$
ranges from $10^8$ to $10^5$. Even for the latter value, the
condition for local equilibrium is not satisfied. This shows
that the breakdown of local equilibrium found here is likely
to occur for typical experimental conditions.

Another way of reaching the same conclusion is the following:
in steady-state growth, the characteristic length scale for the bulk
diffusion field is the diffusion length $l_D=\dbulk/V$. For the
values of $\dbulk$ and $V$ given above, the length scale ranges
from $10^{-9}$ to $10^{-12}$ m, which means that it is comparable
to or even smaller than the typical thickness of an interface.
Furthermore, the time for diffusion across an interface given
by Eq.~(\ref{eq:dperp}) is comparable to or larger than the
time an interface needs to advance by once its thickness.
In solidification, it is well known that {\em solute trapping}
occurs under these conditions, which leads to a breakdown of local
equilibrium. The condition for the occurrence of solute trapping is
\be
\frac{WV}{\dbulk} > 1,
\ee
which is satisfied for the values cited above.

The breakdown of local equilibrium discovered here is thus
very similar to solute trapping in solidification. Our findings 
can therefore be re-stated in a different way: even though surface 
diffusion is faster and controls the growth velocity of the precipitates, 
the quantity that controls local equilibrium at interfaces is the 
low bulk diffusion coefficient. Therefore, non-equilibrium effects 
at interfaces cannot be neglected even at the slow growth velocities 
of discontinuous precipitation.

To avoid confusion, it should be mentioned that of course since
Cahn's work~\cite{Cahn59} it is known that the concentrations
in the {\em volume} of the transformed material are not equal
to the equilibrium concentrations. However, in all existing
theories \cite{Cahn59,Klinger97,Brener99}
it is assumed that the {\em local} concentration in the grain
boundary and the interfaces can be related to the concentration
in the volume of the adjacent growing material by the local
equilibrium assumption.

Since our results show that this assumption is generally not
valid, it is not surprising that none of the existing theories 
agrees with our simulation results. However, Fig.~\ref{fig:abinter}
shows that the system behaves, at least to some degree, as 
predicted by theory, but not with the right value of the
coefficients. This explains why comparisons of experimental data
with theoretical predictions where at least one quantity is
treated as an adjustable parameter can yield reasonable
agreement \cite{Manna01}.

\subsection{Oscillatory motion}

Oscillatory instabilities are well known from cellular and
eutectic solidification in thin samples \cite{Karma96,Ginibre97}.
They occur, like the instability observed here, for spacings
larger than some critical spacing which depends on the alloy
system, the composition, and the processing conditions. The
oscillations are collective, that is, numerous cells or 
lamellae oscillate in phase over a large area of the
solidification front. Since we have only performed here simulations
for a single lamella, it is not clear whether an extended
discontinuous precipitation front exhibits such collective or 
rather an irregular, chaotic behavior. In solidification,
the coupling between neighboring elements is provided by the
diffusion field in the liquid; surface diffusion should provide
a much weaker coupling. Large-scale simulations will be needed
to elucidate this point.

A ``jerky'' or ``stop and go'' motion during discontinuous
precipitation has been observed in several alloy systems, including
Al-Zn \cite{Abdou96}. The oscillatory motion observed in our
simulations could be an explanation for these observations.
More detailed data, both from simulations and experiments,
are needed to clarify this issue.

\section{Conclusion}
\label{sec:conclusion}

We have developed a phase-field model for discontinuous precipitation
and performed simulations to study the influence of various parameters
on the growth velocity of strictly periodic lamellar arrays. Our most
important findings are that (i) for given growth conditions and
composition, steady-state solutions exist for a range of spacings,
(ii) the minimum spacing is given by a limit point beyond which no
steady-state solution exists any more, (iii) an oscillatory
instability occurs for large interlamellar spacings, which leads to
a non-constant growth velocity, and (iv) the breakdown of local equilibrium,
an effect analogous to solute trapping in solidification, cannot be
neglected in discontinuous precipitation.

These results demonstrate that our phase-field model is a promising 
tool to elucidate many open questions on discontinuous precipitation.
Particularly interesting points are the extension of our work to 
lower supersaturations and to growth fronts with several precipitates, 
as well as the further investigation of the oscillatory instability.
To achieve these goals in a reasonable simulation time, more
efficient numerical schemes should be used. In particular,
adaptive meshing seems a promising strategy since the main diffusion
fluxes are concentrated along the interfaces. Another important
question concerns the influence of elastic strains on the
discontinuous precipitation reaction \cite{Brener03}. It is
straightforward to include such effects in multi-phase-field
models \cite{Steinbach06}, but the computational complexity
is dramatically increased, which makes a systematic study 
quite challenging.

Our results also reveal that a fully quantitative modeling of
discontinuous precipitation is a challenging task. As predicted
by Ref.~\cite{Brener99}, the growth velocity depends sensitively
on the angles at the trijunction points and the diffusivity
in the interphase boundary behind the growth front; these parameters
are generally unknown, and difficult to obtain from experiments.
A possible way out would be to use atomistic simulations to obtain
the input parameters for the phase-field model, as already
pioneered in solidification \cite{Hoyt03}.

\bigskip

\noindent{\bf Acknowledgments}

\medskip

L. A. acknowledges financial support through a stipend
by the Minist\`ere de l'En\-seigne\-ment Sup\'erieur et de la 
Recherche Scientifique (Algeria).


\end{document}